\documentclass[sigconf]{acmart}
\usepackage{multirow}
\usepackage{makecell}
\usepackage{bbding}
\usepackage{diagbox}
\usepackage{nicematrix} 
\usepackage{color}
\usepackage{colortbl}
\usepackage{balance} 
\definecolor{skyblue}{RGB}{135,206,235}  

\AtBeginDocument{%
  \providecommand\BibTeX{{%
    \normalfont B\kern-0.5em{\scshape i\kern-0.25em b}\kern-0.8em\TeX}}}

\setcopyright{acmlicensed}
\copyrightyear{2018}
\acmYear{2018}
\acmDOI{XXXXXXX.XXXXXXX}

\acmConference[Conference acronym 'XX]{Make sure to enter the correct
  conference title from your rights confirmation emai}{June 03--05,
  2018}{Woodstock, NY}
\acmISBN{978-1-4503-XXXX-X/18/06}

\copyrightyear{2024}
\acmYear{2024}
\setcopyright{acmlicensed}\acmConference[KDD '24]{Proceedings of the 30th
ACM SIGKDD Conference on Knowledge Discovery and Data Mining}{August
25--29, 2024}{Barcelona, Spain}
\acmBooktitle{Proceedings of the 30th ACM SIGKDD Conference on Knowledge
Discovery and Data Mining (KDD '24), August 25--29, 2024, Barcelona, Spain}
\acmDOI{10.1145/3637528.3671669}
\acmISBN{979-8-4007-0490-1/24/08}
\begin{document}

\title{DIET: Customized Slimming for Incompatible Networks in Sequential Recommendation}


\author{Kairui Fu}
\affiliation{%
  \institution{Zhejiang University}
  \city{Hangzhou}
  \country{China}
}
\email{fukairui.fkr@zju.edu.cn}

\author{Shengyu Zhang}
\authornote{The corresponding author.}
\affiliation{%
  \institution{Zhejiang University}
  \city{Hangzhou}
  \country{China}
}
\affiliation{%
  \institution{Shanghai Institute for Advanced Study of Zhejiang University}
  \city{Shanghai}
  \country{China}
}
\email{sy_zhang@zju.edu.cn}

\author{Zheqi Lv}
\affiliation{%
  \institution{Zhejiang University}
  \city{Hangzhou}
  \country{China}
}
\email{zheqilv@zju.edu.cn}

\author{Jingyuan Chen}
\affiliation{%
  \institution{Zhejiang University}
  \city{Hangzhou}
  \country{China}
}
\email{jingyuanchen@zju.edu.cn}

\author{Jiwei Li}
\affiliation{%
  \institution{Zhejiang University}
  \city{Hangzhou}
  \country{China}
}
\email{jiwei_li@shannonai.com}

\renewcommand{\shortauthors}{Kairui Fu, Shengyu Zhang, Zheqi Lv, Jingyuan Chen, \& Jiwei Li}

\begin{abstract}
    Due to the continuously improving capabilities of mobile edges, recommender systems start to deploy models on edges to alleviate network congestion caused by frequent mobile requests. Several studies have leveraged the proximity of edge-side to real-time data, fine-tuning them to create edge-specific models. Despite their significant progress, these methods require substantial on-edge computational resources and frequent network transfers to keep the model up to date. The former may disrupt other processes on the edge to acquire computational resources, while the latter consumes network bandwidth, leading to a decrease in user satisfaction. In response to these challenges, we propose a customize\textbf{D} sl\textbf{I}mming framework for incompatibl\textbf{E} ne\textbf{T}works(\textbf{DIET}). DIET deploys the same generic backbone (potentially incompatible for a specific edge) to all devices. To minimize frequent bandwidth usage and storage consumption in personalization, DIET tailors specific subnets for each edge based on its past interactions, learning to generate slimming subnets(diets) within incompatible networks for efficient transfer. It also takes the inter-layer relationships into account, empirically reducing inference time while obtaining more suitable diets. We further explore the repeated modules within networks and propose a more storage-efficient framework, DIETING, which utilizes a single layer of parameters to represent the entire network, achieving comparably excellent performance. The experiments across four state-of-the-art datasets and two widely used models demonstrate the superior accuracy in recommendation and efficiency in transmission and storage of our framework.
\end{abstract}



\received{20 February 2007}
\received[revised]{12 March 2009}
\received[accepted]{5 June 2009}

\maketitle

\section{INTRODUCTION}
\label{INTRODUCTION}

With the advent of the era of big data, recommender systems have become indispensable in various aspects of our daily lives, encompassing e-\textit{commerce}, \textit{restaurants}, and \textit{movies}\cite{shisong2023,zhang2023hierarchical,li2023text,liu2023generative,shisong2022,wang2023improving,li2023propensity,li2024noisy,fu2023end,chen2021deep,lv2023parameters}. It is well known that each user frequently interacts with recommender systems, leading to increased latency in adverse network conditions or with large data volumes, due to the necessity of transmitting data back and forth to the cloud for processing. To address this issue, edge-cloud collaborative learning\cite{yao2022edge,mills2021multi,lv2023duet,lv2024intelligent} begins to leverage the advancing capabilities of mobile edges to deploy models directly on edges. This approach allows user \emph{reranking} requests to be processed locally, thereby reducing response latency. Furthermore, various edges have the opportunity to possess distinct models, allowing for differences between cloud and edge models and accommodating both interest and resource heterogeneity.
\begin{figure*}[htb]
    \centering
    \includegraphics[width=1.0\linewidth]{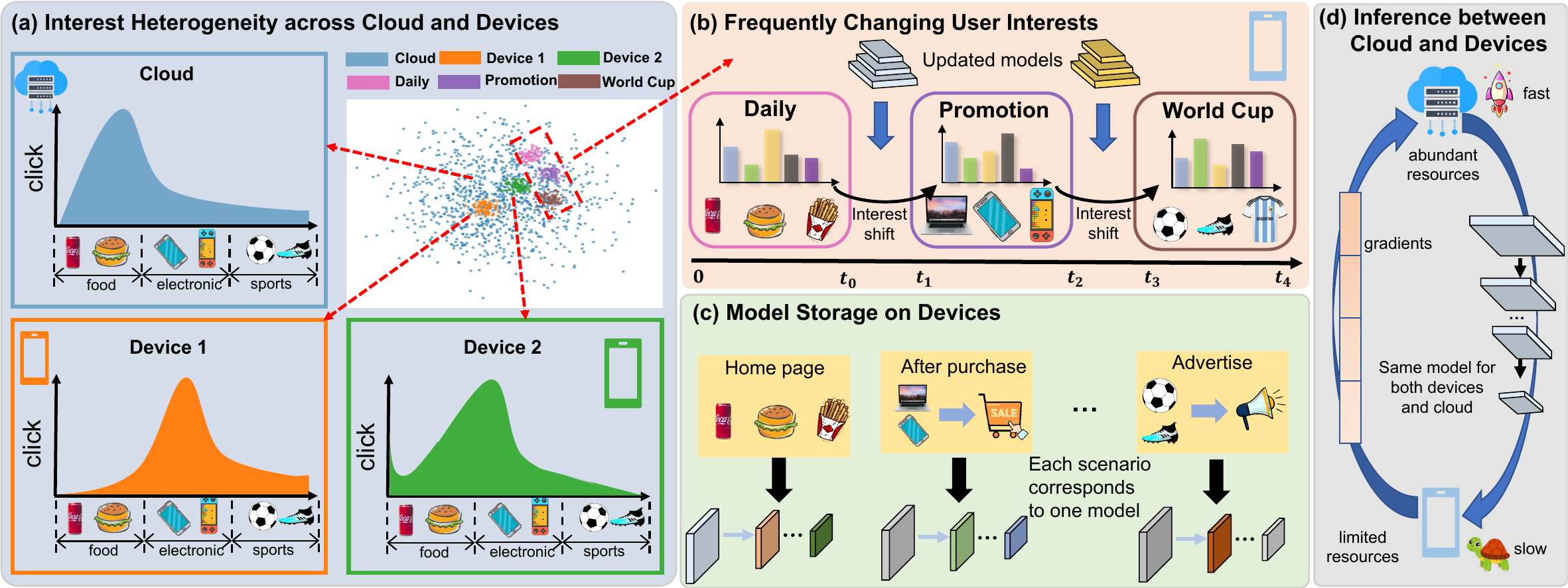}
    \caption{(a): There always exists data distribution shifts across cloud and edges, where cloud holds massive user historical interaction data and each edge only accesses their own data. (b): In each edge, user's interest would change frequently due to some other factors. This would cause cloud to send the latest models to each edge for adaptation, causing massive transmission delays. (c): In recommender systems, different scenarios need different models to provide services, which makes that user devices are flooded with models. (d): Due to the difference in computing resources between cloud and edges, although the model can quickly complete gradient updates and inference on cloud, it still takes a long time on edges.}
    \label{fig:intro}
\end{figure*}

Unfortunately, despite significant advancements in mobile edge capabilities, conducting model training on edges for adaptation of on-edge data still requires a considerable amount of time, and limited data on edges can easily lead to overfitting and performance degradation. In response to this, existing research addresses this issue without introducing much overhead for mobile edges through two perspectives. One approach involves splitting the larger cloud-side model into two parts, deploying the part with a large amount of computation on cloud and lightweight computation on edges\cite{gong2020edgerec,banitalebi2021auto}. Another technique primarily focuses on providing each edge with a personalized model tailored to local data in Figure \ref{fig:intro}. Most of them\cite{lin2020meta,lin2021fr,han2021deeprec, yao2021device} distribute gradient computation across edges, which are then gathered by cloud for aggregation. Despite their advancements, they may result in significant latency due to fine-tuning and transmission delays in order to keep in line with dynamic interests. Other methods\cite{chen2019you,liu2019metapruning,srinivas2017training,frankle2018lottery,lee2018snip,kwon2022fast} alleviate this problem by removing redundant parameters to minimize model size or expedite inference speed. Nevertheless, parameters discovered might be merely redundant and not compatible with some edges. Remaining incompatible parameters lead to poor results for heterogeneous edges, making them degrade a lot.

In light of these considerations, we propose to investigate how to achieve higher performance on different edges under strict resource constraints. Building on this objective, we aim to address three key issues in our investigation: (i) \textbf{Transmission efficiency}. As we discussed above, current methods would download models from cloud frequently to fit user's frequently changing interests in Figure \ref{fig:intro}(b). However, ongoing transmission might lead to severe network congestion and non-negligible transmission delay. Thus, how to design the representation and transmission of the model is a matter worth considering. (ii) \textbf{Storage consumption}. On a superplatform (\textit{e.g.}, Amazon and Taobao), each user has the opportunity to access recommender systems with multiple channels, e.g., \textit{home page recommendation}, \textit{recommendation after purchase}, \textit{short video recommendation}. Under these circumstances, as shown in Figure \ref{fig:intro}(c), previous methods send substantial models to edges, resulting in significant storage pressure. Intuitively, models across different channels are expected to share some similarities, which might help reduce storage occupancy on edges. (iii) \textbf{Inference cost}. The inference speed of edge-side models determines the waiting time after user makes a request. With faster inference speed, recommender system can handle more requests within the same time frame. It presents us with an additional obstacle in decreasing the time required for inference on resource-constrained edges as shown in Figure \ref{fig:intro}(d).

\begin{table}[htb]
  \caption{Brief comparison of DIET and others in edge cloud collaborative learning.} 
  \label{tab:com}
  \centering
  \renewcommand\arraystretch{1.4}
  \resizebox{1.0\columnwidth}{!}{
  
  \begin{NiceTabular}{|c|p{1.73cm}<{\centering}|p{1.73cm}<{\centering}|p{1.73cm}<{\centering}|p{1.73cm}<{\centering}|}
    \toprule[1pt]
    \diagbox{Benefits}{Methods} & \textbf{Base} & \textbf{Fine-tuning} & \textbf{Compress}\footnotemark[1] & \textbf{DIET} \\
     
    \cline{1-5}

    \textbf{Compact} & \XSolidBrush & \XSolidBrush & \Checkmark & \Checkmark \\
    \cline{1-5}

    \textbf{Low Latency} & \XSolidBrush & \XSolidBrush & \Checkmark & \Checkmark \\
    \cline{1-5}

    \textbf{Generalizability} & \XSolidBrush & \Checkmark & \XSolidBrush & \Checkmark \\ 
    \cline{1-5}

    \textbf{Fast Inference} & \XSolidBrush & \XSolidBrush & \Checkmark & \Checkmark \\
    
    \bottomrule[1pt]
    
  \end{NiceTabular}
  }
\end{table}
\footnotetext[1]{It is challenging to achieve two advantages simultaneously through compression.}

Towards this end, we propose a lightweight and efficient edge-cloud collaborative recommendation framework called \textbf{DIET} aiming at customize\textbf{D} sl\textbf{I}mming for incompatibl\textbf{E} ne\textbf{T}works. DIET is dedicated to eliminating incompatible parameters within given networks between cloud and edges while addressing edge-side constraints. Ideally, for distinct edges with various interests, DIET is supposed to assign customized models(including parameters and structures) and minimize costs wherever possible to fit their local interests. Specifically, given each user's real-time sequence, DIET learns to generate personalized \textit{diet} at both the element level and the filter level. A diet consists of a series of binary masks, each of which represents whether the parameter at this location is compatible. In technique, we employ sequence extractor and layer-wise mask generators to extract information from user interactions and learn both element level and filter level importance in the frozen network. Element-level importance makes parameters transmitted between cloud and edges shift from a dense network to lightweight binary masks, reducing the transmission overhead. Filter-level importance not only ensures the exactness of the consolidated learned diets but also accelerates the inference speed empirically. In this situation, for different scenarios, each edge only needs to store one set of parameters, while the cloud can meet the diverse needs of the edge side by training different generators. Once user sends his real-time interactions, cloud will generate the corresponding diet and send it back at once, which prevents user from fine-tuning locally and the time for inference on cloud can be negligible. Motivated by repeated blocks in widely-used recommenders, \textit{e.g.} CNN\cite{tang2018personalized} and transformer\cite{kang2018self}, \textbf{DIETING} is then proposed to use stacked small blocks within the network to represent the entire network, making it more storage-friendly for edges with limited storage. Table \ref{tab:com} provides a brief comparison of \textbf{DIET} and other related methods in edge-cloud collaborative learning, demonstrating the superiority of \textbf{DIET} in edge-cloud collaborative recommendation.

We conduct experiments on four real-world datasets with two widely used recommenders. Further ablation study and parameter analyses consistently validate the generalizability and effectiveness of our approach in edge-cloud collaborative recommendation. We summarize the main contributions of this work as follows:
\begin{itemize}
    \item To the best of our knowledge, we are the first to achieve both structure and parameter personalization under strict edge constraints in edge-cloud collaborative recommendation.
    \item We propose to generate edge-specific diets with user's past interactions, making it fast adaptation of user interests and light-weighted for transfer and storage. 
    \item We account for the importance at filter level to correct the generated diets, which improves the stability of the framework. This approach also empirically reduces the inference cost of edges.
    \item We conduct extensive experiments on four real-world benchmarks, which demonstrate that DIET dramatically achieves better performance under the constraints of limited computational resources and transmission latency on edges.
\end{itemize}

\section{RELATED WORK}
\subsection{Sequential Recommendation}
Owing to the potential to capture dynamic user interests, sequential recommendation has garnered significant attention in both academia and industry. Initial sequential recommender system based on Markov chains\cite{rendle2010factorizing} utilizes the probabilistic modeling of user-item transitions to predict and recommend subsequent items in a sequence. With the flourishing development of deep learning, a series of methods have been applied to recommender system to better capture users’ long and short-term interests, including RNN-based\cite{hidasi2015session,hidasi2016parallel}, transformer-based\cite{kang2018self,zhang2018next,yang2022multi,zhou2018deep,zhang2023adaptive} and CNN-based\cite{tang2018personalized}. Some researchers employed causal-related methods to eliminate bias in recommender systems\cite{zhang2023personalized,zhang2021causerec}. Additionally, some works\cite{liu2023pre, hou2023learning} leveraged transferable knowledge from large language models to enhance the effectiveness of recommender system.

\subsection{Edge-cloud Collobrative Learning}
Recent methods of edge cloud collaborative learning can be categorized into model splitting and model personalization. The former\cite{qian2022intelligent,asheralieva2021auction} divide the model into two parts, which are placed on edges and cloud, thus preventing heavy end-side burden. EdgeRec\cite{gong2020edgerec} deploys the embedding matrix on cloud, while the remaining lightweight model is deployed on user edges. By applying multiple environmental constraints. Auto-Split\cite{banitalebi2021auto} automatically segments each model into cloud-side and edge-side parts, thus avoiding the difficulty of manual segmentation. The latter\cite{ding2020cloud, yao2021device} commit to improving the personality of each edge and their local performance. Detective~\cite{zhang2024revisiting} and ~\cite{chen2024learning} focus on improving the generalization and robustness of cloud models by leveraging multiple distributions and multi-domain data on devices. DCCL\cite{yao2021device} proposes an edge-cloud collaborative framework, introducing a meta-patching mechanism that not only alleviates the computational burden on the edge side but also ensures edge-side personalization. DUET~\cite{lv2023duet} and IntellectReq~\cite{lv2024intelligent} constructe a comprehensive and effective device-cloud collaborative system, respectively addressing the issue of how and when models can be efficiently generalized.

\subsection{Network Compression}
Proper model pruning can help reduce model complexity and computational requirements, improving inference speed and reducing resource consumption. Based on the purpose, model pruning can be categorized into two major types: structured pruning and unstructured pruning. Structured pruning\cite{chen2019you,liu2019metapruning,srinivas2017training} mainly focuses on speeding up inference through removing entire neurons, channels, or layers from the neural network, leading to a more structured model. Although it improves the model’s inference speed, it’s usually challenging to reduce the model size without significantly compromising the final performance. Unstructured pruning\cite{han2015deep} aims to remove individual connections in the neural network, leading to a more irregular reduction of the model size. After LTH\cite{frankle2018lottery} confirmed that there exists a sparse network that can be trained to comparable accuracy in isolation, a variety of methods starts to prune the models during training\cite{sreenivasan2022rare,zhou2019deconstructing,bai2022parameter} or even before training\cite{lee2018snip,kwon2022fast}.

\section{METHOD}
\begin{figure*}[htb]
    \centering
    \includegraphics[width=1.0\linewidth]{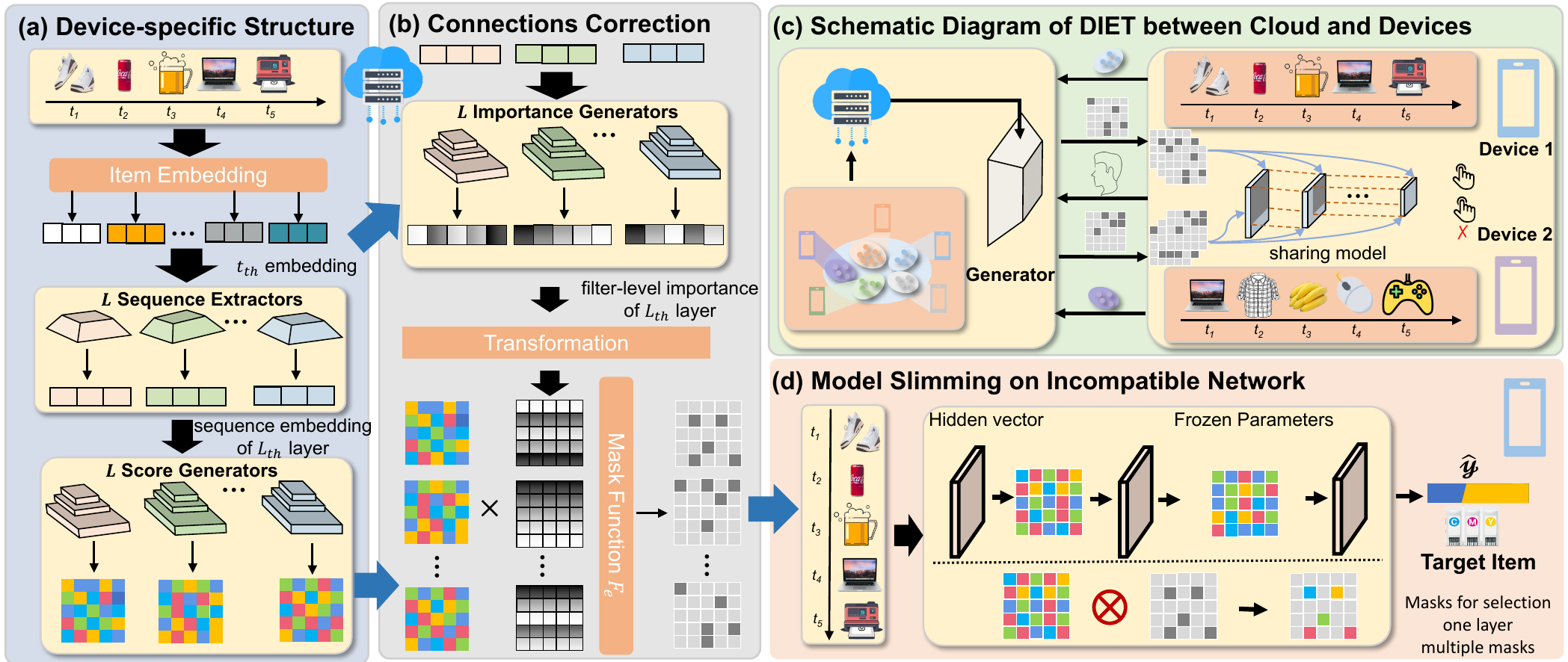}
    \caption{Overview of DIET. (a): Cloud will generate edge-specific diets(subnets) condition on real-time interactions for various data distribution on edges. (b): Another module on cloud to discover inter-layer filter relationships in networks, generated filter-level importance will post-process diets from (a) to correct less important connections. (c): A brief schematic diagram of DIET and DIETING(the portion marked with orange dashed lines). The edges send their real-time samples and cloud customizes diets for them. The transmission over the network consists of a series of binary masks, with all edges sharing the same network. (d): With specific diets from cloud, edges construct their own models, enabling fast adaptation.}
    \label{fig:method}
\end{figure*}

The overall framework of our method is shown in Figure \ref{fig:method}. In this section, DIET is divided into three parts for a clearer introduction: (\romannumeral1) storage saving and efficient transfer in Section \ref{Subnets for Efficient Model Transfer}. (\romannumeral2) edge-specific diets in Section \ref{Edge-specific Diets to Enhance Personality}.  (\romannumeral3) connections correction and light-weighted inference in Section \ref{Link Correction for Each Subnet}.
\subsection{Preliminary}
In the settings of edge-cloud collaborative learning in recommender system, suppose there is an item set $I=[i_1, i_2, ..., i_n]$ and an edge set $D = [d_1, d_2, ..., d_m]$, where $n$ and $m$ are the number of items and edges, separately. Each edge has limited computation resources and cannot access data from other edges. In our paper we rigorously account for model transmission delays and other constraints, aiming to alleviate the issues mentioned in Section \ref{INTRODUCTION}. Apart from this, there exists another cloud server $C$ with abundant computing resources to provide various services for each edge. The cloud server has access to the historical interactions $X^i=[x_1^i, x_2^i, ..., x_T^i]$ of each edge, where $x_t^i \in I$ for $i \in [1,m]$ and $t \in [1,T]$. Each edge also has its own real-time interactions $X^{i^\prime}=[x_1^{i^\prime}, x_2^{i^\prime}, ..., x_t^{i^\prime}]$. Our task in edge-cloud collaborative learning is to improve the generalizability of the cloud model $F$ and further make it adapt to changing environments and resource constraints on edge conditions on their real-time samples:
\begin{equation}
\begin{aligned}
    \mathop{\min}_{\theta} \sum_{i=1}^{i=m}\sum_{j=1}^{j=t}&F(x^i_j|x^i_1, x^i_2, ..., x^i_{j-1};\theta), \\  
    &\mathrm{ s.t. } \mathop{\min} \mathbb{C}.
\end{aligned}
\end{equation}
Here $\mathbb{C}$ represents the set of all edge constraints and $\theta$ denotes the parameters of cloud model $F$.

For a similar consideration, we adopt the local reranking settings as in some mainstream works\cite{lv2023duet,lv2024intelligent,gong2020edgerec} in recommender system. At the beginning of each session or only when user interest changes dramatically, cloud will send the newest models along with the candidate embeddings to each edge. Those embeddings would be stored in local cache for subsequent reranking. Under these circumstances, cloud does not need to participate in edge reranking unless user interests change drastically. As candidate embeddings only occupy a small amount of space, in this paper we mainly consider the storage and transmission of network parameters.

\subsection{Model Slimming for Incompatible Parameters}
\label{Subnets for Efficient Model Transfer}
Network bandwidth and storage resources for the edges continue to be in short supply despite the rapid development of technology. The former one makes it more expensive for the edges to acquire the latest model from cloud. The latter one may become some obstacles where each edge needs to keep multiple models for various scenarios, thereby adding burden to itself. This situation is especially obvious in recommendation systems, where the user’s interests are changing frequently and each user may interact with recommenders through various ways in some large e-commerce system. Fortunately, previous work\cite{zhou2019deconstructing} has proved the potential of random-initialized networks, i.e. where there is a sub-network that can achieve comparable performance through parameter selection as training the entire network. Inspired by this, we propose to discover whether this strategy can be effective in recommender system.

Consider that the proposed model $F_{P}$ with $K$ transformers, each with parameters $W_i=[w_i^0, w_i^1, ..., w_i^{p-1}]$, where $i \in [0,K)$ and $p$ is the number of linear layers in $i$th transformer. For those transformers, the weights $W_i$ are frozen and the training process can be converted to find an optim mask $M_i=[m_i^0, m_i^1, ..., m_i^{p-1}]$ for each linear layer. Then the weight of each linear layer for inference will be:
\begin{equation}
\label{eq:1}
    W_i^{\prime} = [w_i^0 \odot m_i^0, w_i^1 \odot m_i^1, ..., w_i^{p-1} \odot m_i^{p-1}].
\end{equation}

In pursuit of this, for each linear layer, we assign another parameter $S_i=[s_i^0, s_i^1, ..., s_i^p]$ to indicate the importance score of each element in this layer. Then Equation \ref{eq:1} can be written as:
\begin{equation}
\label{eq:2}
    W_i^{\prime} = [w_i^0 \odot F^{\prime}(s_i^0), w_i^1 \odot F^{\prime}(s_i^1), ..., w_i^{p-1} \odot F^{\prime}(s_i^{p-1})],
\end{equation}
where $F^{\prime}$ is a specific function used to extract $m_i^0$ from $s_i^0$. Given the hyperparameter $\alpha$, the output of $F^{\prime}$ will be 1 if the absolute value of the element is among the top $\alpha \%$ absolute values in $s_i^0$, otherwise the output is 0. To overcome the non-differential problem of $F^{\prime}$, we choose straight through estimator rather than performing sampling on continuous variable $S_i$ to prevent extra sampling process:
\begin{equation}
\label{eq:3}
    \frac{\partial L}{\partial s_i^0} \approx \frac{\partial L}{\partial W_i^{\prime}} \frac{\partial W_i^{\prime}}{\partial F^{\prime}(s_i^0)}.
\end{equation}

Since current networks are mainly composed of stacked blocks, we begin to conjecture whether the initialization values in the network are important. In light of this, we propose another variant of DIET named \textbf{DIETING}, which is more lightweight than DIET. \textbf{DIETING} proposes that since the network consists of identical blocks, it might be also feasible to initialize the parameters of each layer and use one of the layers to initialize the entire network. That is, for all $i \in K$ and $j \in P$,
\begin{equation}
    w_i^0 = W_{max}[:size(w_i^0)],
\end{equation}
where $W_{max}$ contains the most parameters among all layers. Under these circumstances, all models can be represented with one layer and a series of masks. This not only reduces the storage overhead but also the transmission delay because for cloud the transferred parameters become binary form, reducing the size of transmission.

\subsection{Edge-specific Diets to Enhance Personality}
\label{Edge-specific Diets to Enhance Personality}
Although sending binary masks between cloud and edges reduces the latency required for transmission, fewer meaningful parameters on edges may reduce the model’s generalizability, as parameters suitable for cloud may not necessarily be suitable for edges. Motivated by the widely used hypernetwork\cite{ha2017hypernetworks,alaluf2022hyperstyle,dinh2022hyperinverter}, which generates the corresponding parameters with learnable vectors, this capability can be integrated into our framework to produce personalized diets for each edge. Formally, the hypernetwork $G$ get an input $z$, then output the specific weight $w_G$:
\begin{equation}
\label{eq:4}
    w_G = G(z).
\end{equation}

Despite the personalized parameters it produces, the hypernetwork uses a random latent vector $z$ as its input, which needs to be retrained for each edge on their local data. However, for resource-constrained edges, finetuning the model on edges to get distinctive $z$ is not feasible. We thereby introduce an extra sequence extractor to learn to generate the latent vector $z$ from user's recent behaviors ourselves. Each edge uploads its recent interactions and cloud can extract valuable information from them to generate personalized diets. Then Equation \ref{eq:4} becomes:
\begin{equation}
\label{eq:5}
    w_G = G(E(X)),
\end{equation}
where $X$ is the real-time interactions of each edge.

\subsubsection{Sequence Extractor}
Given the length of the interaction as $l$ and its representation $g\in R^{l\times d}$, with $d$ being the dimension of the item embedding, the sequence extractor aims to extract those underlying information $g^{\prime} \in R^{d}$ representing the data distribution of each edge. Though most of the DNN architectures can be used as the sequence extractor, in our framework we use GRU\cite{chung2014empirical} as the sequence extractor which offers efficient training, reduced complexity, and comparable performance to LSTMs in capturing long-range dependencies. In order to prevent mutual interference between different layers, each layer has its own independent extractor when calculating its unique mask.

\subsubsection{HyperNetwork} In this section, we will outline how to utilize the extracted sequence features $g^{\prime}$ and hypernetwork $G$ to generate diet for each layer. Take the linear layer as an example, the dimension of it is $d_{out} \times d_{in}$, with $d_{out}$ and $d_{in}$ are the dimension of its output and input, respectively. The hypernetwork can consist of a fully connected layer or a complex MLP.  Similarly, every layer has its hypernetwork to ensure an adequate level of parameter personalization.

The generated score will then replace the importance score $S_i$ to produce the corresponding mask $W_i^{\prime}$ as in Equation \ref{eq:2}:
\begin{equation}
\label{eq:6}
    W_i^{\prime} = [w_i^0 \odot F^{\prime}(G_i^0(E_i^0(X))), ... , w_i^{p-1} \odot F^{\prime}(G_i^{p-1}(E_i^{p-1}(X)))],
\end{equation}
where $G_i^{p-1}$ and $E_i^{p-1}$ are the hypernetwork and sequence extractor of the $p-1$th linear layer in the $i$th transfomer.

Despite extra model parameters we include to generate masks, those modules won't be sent to each edge but rather kept in cloud. Hence it won't add overhead for edges. Once users upload their real-time samples $X$ to the cloud, the sequence extractors and hypernetworks of each layer will generates the personalized diets according to the extracted information of real-time data samples and send them to the edge. Multiple extractors and hypernetworks can be executed in parallel during an inference process, thus incurring minimal waiting time overhead.

\subsection{Connections Correction for Each Diet}
\label{Link Correction for Each Subnet}
Recall that when generating masks in the previous steps, each element in the linear layer is independently predicted and only element-level dependencies are considered. Inspired by previous work\cite{ganesh2021mint} which delved into the inter-layer filter relationships and emphasized the retention of filters containing substantial information across adjacent layers, we propose to leverage inter-layer dependencies to capture more comprehensive information and enhance the overall model performance.

Similarly, we take the hypernetwork mentioned in Section \ref{Edge-specific Diets to Enhance Personality} to generate inter-layer importance. To simplify the architecture, the element-level sequence extractor is repurposed for the filter level. In our view, those extracted common characteristics can be used for both importance calculations. Formally, given the extracted features $g^{\prime}$, another hypernetwork $G^{\prime}_i$ maps it to the row level importance $r_i \in R^{d_{out}}$, which will be expanded and multiply the generated diets:
\begin{equation}
    w_i^{0\prime}[a,b] = w_i^0[a,b] \times F^{\prime}(G_i^0(E_i^0(X)[a,b] \times r_i^0[a]),
\end{equation}
where
\begin{equation}
    r_i^0 = \mathop{softmax}(G^{0{\prime}}_i(E_i^0(X))),
\end{equation}
the \textit{softmax} function here is to prevent some outputs from being negative because the mask generate function $F^{\prime}$ will use the absolute value to determine whether to retain an element at a certain position.

Taking inter-layer dependencies into consideration not only provides a better diet for edges but also empirically speeds up the inference on each edge. The results will be described in Section \ref{Overall Performance}. Given the varying importance of each row/filter within a layer, elements in less important filters will receive lower scores. This prompts the importance of the elements within this filter to be positioned towards the latter part of the entire layer, ultimately resulting in the entire filter not being selected and therefore not needing to participate in the inference calculation, thereby reducing the inference overhead on edges.
\begin{table}[h]
  \caption{Statistics of the datasets.}
  \label{dataset}
  \centering
  \resizebox{1.0\columnwidth}{!}{
  \begin{tabular}{lccccc}
    \toprule
    Dataset     & \#Users     & \#Items     & \#Interactions     & \#SeqLen     & \#Sparsity \\
    \midrule
    ML-1M & 6,040  & 3,012  & 994,852  & 164.71  & 94.52\%     \\
    ML-100K & 943  & 928  & 94,672  & 100.45  & 89.16\%     \\
    CD     & 78,318  & 57,326  & 1,955,164  & 24.96  & 99.96\%     \\
    TV     & 31,482  & 68,308  & 867,853  & 27.56  & 99.96\%     \\
    \bottomrule
  \end{tabular}
  }
\end{table}

\begin{table*}[htb]
  \caption{Overall Performance on recommendation and resources. We use bold font to denote the best model and underline the next best-performing model. $\uparrow$ and $\downarrow$ denote that larger and smaller metrics lead to better performance, respectively.} 
\vspace{-0.3cm}
\label{tab:experiment_rec}
\centering
\resizebox{1.0\textwidth}{!}{
    \begin{tabular}{c|c|p{1.15cm}<{\centering}|p{1.15cm}<{\centering}|p{1.15cm}<{\centering}|p{1.15cm}<{\centering}|p{1.15cm}<{\centering}|p{1.15cm}<{\centering}|p{1.15cm}<{\centering}|p{1.15cm}<{\centering}|p{1.15cm}<{\centering}|p{1.15cm}<{\centering}|p{1.15cm}<{\centering}|p{1.15cm}<{\centering}|p{1.15cm}<{\centering}|p{1.15cm}<{\centering}|p{1.15cm}<{\centering}|p{1.15cm}<{\centering}}
    \toprule[2pt]

     \multirow{3}{*}{\textbf{Model}} & \multirow{3}{*}{\textbf{Method}} & \multicolumn{16}{c}{\textbf{Dataset}} \\
     \cline{3-18}
     & & \multicolumn{4}{c|}{\textbf{MovieLens-1M}} & \multicolumn{4}{c|}{\textbf{MovieLens-100k}} & \multicolumn{4}{c|}{\textbf{Amazon-CD}} & \multicolumn{4}{c}{\textbf{Amazon-TV}} \\
     \cline{3-18}
     & & \textbf{NDCG $\uparrow$} & \textbf{Hit $\uparrow$} & \textbf{Param $\downarrow$} & \textbf{FLOPs $\downarrow$} & \textbf{NDCG $\uparrow$} & \textbf{Hit $\uparrow$} & \textbf{Param $\downarrow$} & \textbf{FLOPs$\downarrow$} & \textbf{NDCG $\uparrow$} & \textbf{Hit $\uparrow$} & \textbf{Param $\downarrow$} & \textbf{FLOPs $\downarrow$} & \textbf{NDCG $\uparrow$} & \textbf{Hit $\uparrow$} & \textbf{Param $\downarrow$} & \textbf{FLOPs $\downarrow$} \\
     \bottomrule[1pt]
     \bottomrule[1pt]
     
     \multirow{10}{*}{\textbf{SASRec}} & \texttt{Base} & \underline{0.0974}  & 0.1849 & 1.3107 & 0.2086 & 0.0517 & 0.1077 & 1.3107 & 0.2086 & 0.0386 & 0.0529 & 1.3107 & 0.2086 & 0.0665 & 0.0835 & 1.3107 & 0.2086 \\
     & \texttt{Random} & 0.0926 & 0.1836 & 0.7859 & 0.2086 & 0.0569 & 0.1265 & 0.2618 & 0.2086 & 0.0409 & 0.0557 & 0.2618 & 0.2082 & 0.0694 & 0.0871 & 0.2618 & 0.2086 \\
     & \texttt{LTH} & 0.0962 & 0.1852 & 0.7859 & 0.2086 & 0.0564 & 0.1222 & 0.2618 & 0.2086 & 0.0408 & 0.0564 & 0.2618 & 0.2086 & \underline{0.0703} & \underline{0.0891} & 0.2618 & 0.2082 \\
     & \texttt{SNIP} & 0.0932 & 0.1835 & 0.7860 & 0.2073 & \underline{0.0582} & 0.1266 & 0.2620 & 0.1999 & 0.0405 & 0.0545 & 0.2620 & 0.1439 & 0.0699 & 0.0875 & 0.2618 & 0.1375 \\
     & \texttt{Rare Gem} & 0.0949 & 0.1868 & 0.7866 & 0.2086 & 0.0578 & 0.1243 & 0.2624 & 0.2044 & \underline{0.0421} & \underline{0.0582} & 0.2624 & 0.1961 & 0.0696 & 0.0883 & 0.2624 & 0.1919  \\
     & \texttt{Supermask} & 0.0723 & 0.1505 & \textbf{0.0410} & 0.2086 & 0.0518 & 0.1147 & \textbf{0.0410} & 0.1135 & 0.0301 & 0.0398 & \textbf{0.0410} & 0.1580 & 0.0574 & 0.0701 & \textbf{0.0410} & 0.1532  \\
     & \texttt{PEMN} & 0.0859 & 0.1769 & \textbf{0.0410} & 0.2078 & 0.0538 & 0.1239 & \textbf{0.0410} & 0.1743 & 0.0408 & 0.0554 & \textbf{0.0410} & 0.1769 & 0.0687 & 0.0863 & \textbf{0.0410} & 0.1916  \\
     & \texttt{Gater} & 0.0963 & \underline{0.1886} & 1.4459 & \underline{0.1167} & 0.0555 & 0.1177 & 0.8110 & \underline{0.0671} & 0.0412 & 0.0568 & 1.2329 & \underline{0.1197} & 0.0701 & 0.0888 & 0.9544 & \underline{0.0783}\\
     & \texttt{STTD} & 0.0863 & 0.1714 & 0.0461 & 0.3110 & \underline{0.0582} & \underline{0.1283} & 0.0461 & 0.3110 & 0.0385 & 0.0503 & 0.0461 & 0.3110 & 0.0692 & 0.0858 & 0.0461 & 0.3110 \\
     \cline{2-18}
     \rowcolor{gray!40}
    \cellcolor{white} & \texttt{DIET} & \textbf{0.1008} & \textbf{0.1929} & \textbf{0.0410} & \textbf{0.1022} & \textbf{0.0635} & \textbf{0.1319} & \textbf{0.0410} & \textbf{0.0416} & \textbf{0.0425} & \textbf{0.0590} & \textbf{0.0410} & \textbf{0.1154} & \textbf{0.0707} & \textbf{0.0896} & \textbf{0.0410} & \textbf{0.0764} \\
     \rowcolor{skyblue!70}
    \cellcolor{white} & \texttt{Improv} & 3.49\% & 4.33\% & $\times$ 31.97 & $\times$ 2.04 & 22.82\% & 22.47\% & $\times$ 31.97 & $\times$ 5.01 & 10.96\% & 11.53\% & $\times$ 31.97 & $\times$ 1.81 & 6.32\% & 7.31\% & $\times$ 31.97 & $\times$ 2.73 \\
    
    \midrule[1pt]
    \midrule[1pt]
    
    \multirow{10}{*}{\textbf{Caser}} & \texttt{Base} & 0.0984 & 0.1820 & 0.4922 & 0.0586 & 0.0518 & 0.1065 & 0.4922 & 0.0586 & 0.0310 & 0.0424 & 0.4922 & 0.0586 & 0.0569 & 0.0719 & 0.4922 & 0.0586 \\
    & \texttt{Random} & 0.0826 & 0.1565 & 0.0983 & 0.0580 & 0.0483 & 0.1001 & 0.0983 & 0.0580 & 0.0251 & 0.0363 & 0.1967 & 0.0580 & 0.0468 & 0.0606 & 0.0983 & 0.0580  \\
    & \texttt{LTH} & 0.0959 & 0.1759 & 0.0986 & 0.0580 & 0.0477 & 0.0973 & 0.0986 & 0.0570 & \underline{0.0322} & 0.0424 & 0.1971 & 0.0583 & \underline{0.0586} & \underline{0.0724} & 0.0519 & 0.0986  \\
    & \texttt{SNIP} & 0.0904 & 0.1712 & 0.0986 & 0.0577 & 0.0467 & 0.0968 & 0.0986 & 0.0577 & 0.0298 & 0.0405 & 0.1971 & 0.0583 & 0.0497 & 0.0634 & 0.0986 & 0.0572  \\
    & \texttt{Rare Gem} & 0.0953 & 0.1757 & 0.0986 & 0.0577 & 0.0509 & 0.1067 & 0.0986 & 0.5523 & 0.0308 & \underline{0.0426} & 0.1970 & 0.0580 & 0.0537 & 0.0676 & 0.0986 & 0.0561  \\
    & \texttt{Supermask} & 0.0797 & 0.1572 & \textbf{0.0154} & 0.0586 & 0.0360 & 0.0797 & \textbf{0.0154} & 0.0578 & 0.0213 & 0.0318 & \textbf{0.0154} & 0.0586 & 0.0403 & 0.0522 & \textbf{0.0154} & 0.0576  \\
    & \texttt{PEMN} & 0.0781 & 0.1495 & \textbf{0.0154} & 0.0364 & 0.0371 & 0.0812 & \textbf{0.0154} & 0.0552 & 0.0213 & 0.0316 & \textbf{0.0154} & 0.0544 & 0.0358 & 0.0487 & \textbf{0.0154} & 0.0577  \\
    & \texttt{Gater} & 0.0837 & 0.1573 & 0.3891 & \underline{0.0298} & 0.0541 & 0.1060 & 0.6929 & \underline{0.0456} & 0.0257 & 0.0358 & 0.3400 & \underline{0.0343} & 0.0537 & 0.0676 & 0.5983 & \underline{0.0496}  \\
    & \texttt{STTD} & \textbf{0.0989} & \textbf{0.1826} & 0.0231 & 0.0996 & \underline{0.0566} & \underline{0.1103} & 0.0231 & 0.0996 & 0.0270 & 0.0360 & 0.0231 & 0.0996 & 0.0555 & 0.0697 & 0.0231 & 0.0996 \\
    \cline{2-18}
    \rowcolor{gray!40}
    \cellcolor{white} & \texttt{DIET} & \underline{0.0987} & \underline{0.1821} & \textbf{0.0154} & \textbf{0.0280} & \textbf{0.0572} & \textbf{0.1143} & \textbf{0.0154} & \textbf{0.0333} & \textbf{0.0356} & \textbf{0.0488} & \textbf{0.0154} & \textbf{0.0294} & \textbf{0.0617} & \textbf{0.0771} & \textbf{0.0154} & \textbf{0.0488} \\
    \rowcolor{skyblue!70}
    \cellcolor{white} & \texttt{Improv} & 0.31\% & 0.06\% & $\times$ 31.96 & $\times$ 2.09 & 10.42\% & 7.32\% & $\times$ 31.96 & $\times$ 1.76 & 14.84\% & 15.09\% & $\times$ 31.96 & $\times$ 1.99 & 8.44\% & 7.23\% & $\times$ 31.96 & $\times$ 1.20 \\

    \bottomrule[2pt]
    \end{tabular}
}
\end{table*}
\section{EXPERIMENTS}
\label{EXPERIMENTS}
In order to demonstrate the effectiveness and efficiency of our method, we compare it with other outstanding methods on two most widely used models and four real-world recommendation datasets. We choose \textit{SASRec} and \textit{Caser} as the architectures they use are CNN and transformer respectively. \textbf{All results are the average of five experiments with five individual random seeds}.

\subsection{Experimental Setup}

\subsubsection{Datasets}
The experiments are conducted on four state-of-art benchmarks \emph{Movielens-1M}\footnote{\url{https://grouplens.org/datasets/movielens/1m}}, \emph{Movielens-100K}\footnote{\url{https://grouplens.org/datasets/movielens/100k}}, \emph{Amazon-CD} and \emph{Amazon-TV}\footnote{\url{https://nijianmo.github.io/amazon/index.html}}. Detailed statistics of them are shown in Table \ref{dataset}. To keep data quality, we use 20-core setting for Movielens datasets and 10-core settings for Amazon datasets according to different data sparsity of each dataset. We treat the interactions with positive ratings as positive samples. Following previous works, we sort the user-item interactions in chronological order and treat the last interacted item of each user as test sample.

\subsubsection{Evaluation Metrics}
For the edge-cloud collaborative learning in recommendation, we are supposed to consider recommender performance, transmission latency, and inference speed simultaneously. For the recommender performance, we use Hit and NDCG, two frequently used metrics in recommender system. For the transmission latency, we evaluate it with the number of bits needed to transfer when updating. For the inference speed, we use Floating-point operations (FLOPs) of a single iteration during inference. Detailed descriptions can be found in Appendix.

\subsubsection{Base model \& Baselines}
To verify the universality of our method towards both convolutional layer and linear layer, we choose the following two widely used sequential recommenders:

\begin{itemize}
\item {\textbf{SASRec}} is a transformer-based sequential recommender, it leverages the attention mechanism to capture user's interests and assign weights for each item towards the target item.
\item {\textbf{Caser}} regards each input interaction as picture and makes use of horizontal and vertical convolutional layers to capture both point-level and union-level interests of each user.
\end{itemize}

We compare our method with a variety of compression-based methods, of which the descriptions are shown as below:
\begin{itemize}
    \item {\textbf{Base}} simply trains the sequential recommender on cloud and deloys it on edge without any extra training.
    \item {\textbf{Random}} randomly masks weights in each layer then train the sparse model from scratch. 
    \item {\textbf{LTH}}\cite{frankle2018lottery} tends to find a \emph{lottery ticket} from a randomly initialized network, which can achieve comparable performance after training for at most the same number of iterations.
    \item {\textbf{SNIP}}\cite{lee2018snip} chooses to selectively prune redundant connections based on their sensitivity even before any training.
    \item {\textbf{GEM-MINER}}\cite{sreenivasan2022rare} suggests that subnet with better accuracy at initialization can achieve better accuracy after training.
    \item {\textbf{SuperMask}}\cite{zhou2019deconstructing} first discover that there exists a subnetwork from a randomly initialized network that can get similar performance without any training.
    \item {\textbf{PEMN}}\cite{bai2022parameter} further validate the potential of random weight by learning various masks.
    \item {\textbf{Gater}}\cite{chen2019you} aims to get rid of unnecessary calculations and acquire higher inference speed.
    \item {\textbf{STTD}}\cite{xia2022device} represents each layer with small matrices through multiplication to reduce parameters.
\end{itemize}

\subsection{Overall Performance}
\label{Overall Performance}
The comparison of the recommendation performance of our method and other baselines on four datasets is shown in Table \ref{tab:experiment_rec}. From the table, it is evident that most of the pruning methods perform worse than the basic recommender without any fine-tuning, indicating that the \emph{lottery ticket} found by them does not have a guarantee of comparable performance with base recommender. Surprisingly, those data-dependent pruning methods sometimes perform slightly worse than random pruning. This could be attributed to the rapid changes in user interests within recommender systems, leading to a distribution gap between real-time data, rendering the sparse network learned through training data ineffective.

In a nutshell, our approach consistently outperforms the base recommender and other baselines across two models, four datasets, and two evaluation metrics. Although it does not exhibit significant improvement over Caser on Movielens-1M, it still yields the best performance among all the other baselines, which drops considerably compared to the original recommender. Moreover, DIET improves the base recommender by a large margin on various datasets. For example, on SASRec-based Movielens-100K, we achieve almost $20\%$ improvement over the base recommender and on Amazon-CD dataset, we improve the two models by more than $10\%$ concerning NDCG and Hit through searching the specific diets.

Simply learning from given networks is far from enough and it might cause inferior results. Those sparse selection-based methods, like \emph{SuperMask} and \emph{PEMN}, become the two worst performing methods. The heterogeneity of interests poses more challenges to those methods, making them less likely to find the proper mask. In contrast, \textbf{DIET} confronts the heterogeneity of user behaviors by generating corrected personalized diet for each edge given its recent interactions.

We also present the number of parameters to be transferred and FLOPs for inference relative to the base recommender. According to the result, we can find that all of the pruning methods reduce the model size a lot, among which those sparse selection-based methods get more significant effects, reducing to $3\%$ of the origin model size. Despite the substantial achievement in model compression, these methods do not improve the inference time, as the model still takes about the same amount of time for inference. While \emph{Gater} largely reduces the inference time, it needs to transfer a large number of parameters and perform poorly on some datasets.  Although \emph{STTD} achieve better results on Movielens-1M of \emph{Caser}, it degrades a lot compared to even the base model on other datasets. Moreover, due to the additional computation required for the semi-tensor product’s matrix multiplication, \emph{STTD} requires more computational resources for each inference. In comparison, DIET shows consistent improvement on both transmission and inference, especially achieves the best results over Caser on Amazon-CD. All the above results demonstrate the capability and efficiency of DIET in edge-cloud collaborative learning.

\begin{table}[htb]
\caption{Ablation Study} 
\vspace{-0.3cm}
\label{tab:experiment_ablation}
\centering
\resizebox{1.0\linewidth}{!}{
    \begin{tabular}{c|c|p{1.0cm}<{\centering}|p{1.0cm}<{\centering}|p{1.0cm}<{\centering}|p{1.0cm}<{\centering}|p{1.0cm}<{\centering}|p{1.0cm}<{\centering}|p{1.0cm}<{\centering}|p{1.0cm}<{\centering}}
    \toprule[2pt]
     \multirow{3}{*}{\textbf{Dataset}} & \multirow{3}{*}{\textbf{Metric}} & \multicolumn{8}{c}{\textbf{Method}} \\
     \cline{3-10} 
     & & \multicolumn{4}{c|}{\textbf{SASRec-based}} & \multicolumn{4}{c}{\textbf{Caser-based}} \\
     \cline{3-10} 
     & & \textbf{Base} & \textbf{+mask} & \textbf{+MG} & \textbf{DIET} & \textbf{Base} & \textbf{+mask} & \textbf{+MG} & \textbf{DIET}\\
     
    \midrule[1pt]
    \midrule[1pt]
    \multirow{4}{*}{\textbf{Movielens-1M}} & \textbf{NDCG$@10$} & 0.0974 & 0.0859 & 0.0990 & 0.1009 & 0.0984 & 0.0781 & 0.0930 & 0.0987\\
    & \textbf{Hit$@10$} & 0.1849 & 0.1769 & 0.1898 & 0.1929 & 0.1820 & 0.1495 & 0.1744 & 0.1821\\
    & \textbf{Params} & 1.3107 & 0.0410 & 0.0410 & 0.0410 & 0.4922 & 0.0154 & 0.0154 & 0.0154\\
    & \textbf{FLOPs} &  0.2086 & 0.2078 & 0.2078 & 0.1022 & 0.0586 & 0.0364 & 0.0364 & 0.0280\\
    \cline{1-10}
    \multirow{4}{*}{\textbf{Movielens-100K}} & \textbf{NDCG$@10$} & 0.0517 & 0.0538 & 0.0631 & 0.0635 & 0.0518 & 0.0371 & 0.0546 & 0.0572 \\
    & \textbf{Hit$@10$} & 0.1077 & 0.1239 & 0.1313 & 0.1319 & 0.1064 & 0.0812 & 0.1080 & 0.1143\\
    & \textbf{Params} & 1.3107 & 0.0410 & 0.0410 & 0.0410 & 0.4922 & 0.0154 & 0.0154 & 0.0154\\
    & \textbf{FLOPs} &  0.2086 & 0.1743 & 0.1743 & 0.0416 & 0.0586 & 0.0552 & 0.0552 & 0.0333\\
    \cline{1-10}
    \multirow{4}{*}{\textbf{Amazon-CD}} & \textbf{NDCG$@10$} & 0.0386 & 0.0408 & 0.0409 & 0.0425 & 0.0310 & 0.0213 & 0.0357 & 0.0356\\
    & \textbf{Hit$@10$} & 0.0529 & 0.0554 & 0.0553 & 0.0590 &0.0424 & 0.0316 & 0.0485 & 0.0488 \\
    & \textbf{Params} & 1.3107 & 0.0410 & 0.0410 & 0.0410 & 0.4922 & 0.0154 & 0.0154 & 0.0154 \\
    & \textbf{FLOPs} & 0.2086 & 0.1769 & 0.1769 & 0.0815 & 0.0586 & 0.0544 & 0.0544 & 0.0294\\

    \bottomrule[2pt]
    \end{tabular}
}
\end{table}

\subsection{Ablation Study}
\label{Ablation Study}
After having a full comparison between our method and others, we would like to learn more about its details and check whether each part of this design has played its intended role. To achieve this, we incrementally incorporate various components into \textbf{DIET} and analyze the impact of the resulting models individually. The ablated models are presented below, and results are depicted in Table \ref{tab:experiment_ablation}:

\begin{itemize}
    \item {\textbf{Base}} are the two origin sequential recommenders SASRec and Caser. Both of them exhibit similar performance on Movielens-1M and Movielens-100K. However, we notice that on Amazon-CD SASRec gains more NDCG and Hit improvements against Caser. This owns to the more parameters of SASRec to capture high-order user interests.
    \item {\textbf{+mask}} neither use hypernetwork to generate personalized mask for each user nor correct the mask. This means that we will generate the same mask for all users. From the table we can observe a clear performance drop of \emph{+mask} over Caser on all three datasets even if it yields better performance over Movielens-100K and Amazon-CD. We attribute this to the complex and ever-changing user interests in recommender system as mentioned in Section \ref{Overall Performance}. Not surprisingly, by employing lightweight diets, the number of parameters requiring transfer is significantly reduced, resulting in a substantial decrease in communication costs.
    \item {\textbf{+MG}} generate personalized masks based on the past few interactions but it does not consider filter-level importance and treats each connection as independent. This model transmits the same number of parameters as \emph{+mask} due to the same sparse selection strategy. What's more, we observe a huge improvement on recommendation performance over almost all datasets. This underscores the effectiveness of the personalized mask generation method based on user interests. Nevertheless, introducing hypernetwork does not change the number of FLOPs during inference and it obtains similar results and poorer results to \emph{+mask} and \emph{Base} models individually.
    \item {\textbf{DIET}} aims to rectify the generated mask by taking the filter-level importance into consideration, thereby creating interdependencies among elements. Compared to the \emph{+mask} model, \emph{DIET} superior results on all metrics. First, owning to the corrected mask, the recommendation performance improves greatly and becomes the best among all the settings. Additionally, by reducing the weights of less important filters, more filters contain elements that are all zero, leading to a substantial reduction in FLOPs and accelerated inference on edges. The number of Floating Point Operations has dropped by at least $50\%$ from the original basis. These findings once again confirm the effectiveness of our framework in edge-cloud recommendation.
\end{itemize}

\subsection{Influence of DIETING}
In this section, we aim to illustrate within our framework that the connections learned during the training process take precedence over the initial values. Through experiments conducted on the aforementioned datasets, the results are presented in Table \ref{tab:DIETING}. Surprisingly, DIETING, which initializes the model with just one layer, achieves comparable performance to DIET. This further confirms that the initial parameters of the model have minimal impact on the final results. Instead, the crucial factor lies in the customization of diets for each edge. The composition of elements within the neural network plays a pivotal role in determining the ultimate outcome. With DIETING, the parameters stored on edges can consist of only one layer, regardless of the scenario.
    
\begin{table}[htb]
\caption{Overall comparison between DIETING and DIET} 
\vspace{-0.3cm}
\label{tab:DIETING}
\centering
\resizebox{1.0\linewidth}{!}{
    \begin{tabular}{c|c|p{1.0cm}<{\centering}|p{1.0cm}<{\centering}|p{1.0cm}<{\centering}|p{1.0cm}<{\centering}|p{1.0cm}<{\centering}|p{1.0cm}<{\centering}|p{1.0cm}<{\centering}|p{1.0cm}<{\centering}}
    \toprule[2pt]
     \multirow{3}{*}{\textbf{Dataset}} & \multirow{3}{*}{\textbf{Method}} & \multicolumn{8}{c}{\textbf{Metric}} \\
     \cline{3-10}
     & & \multicolumn{4}{c|}{\textbf{SASRec-based}} & \multicolumn{4}{c}{\textbf{Caser-based}} \\
     \cline{3-10}
     & & \textbf{NDCG} & \textbf{Hit} & \textbf{Params} & \textbf{FLOPs} & \textbf{NDCG} & \textbf{Hit} & \textbf{Params} & \textbf{FLOPs} \\
     \cline{1-10}
     \multirow{2}{*}{\textbf{MovieLens-1M}} & \textbf{DIET} & 0.1008 & 0.1929 & 0.0410 & 0.1022 & 0.0987 & 0.1821 & 0.0154 & 0.0280 \\
     & \textbf{DIETING} & 0.1005 & 0.1919 & 0.0410 & 0.0726 & 0.0974 & 0.1791 & 0.0154 & 0.0273 \\
     \midrule[1pt]
     \multirow{2}{*}{\textbf{MovieLens-100K}} & \textbf{DIET} & 0.0635 & 0.1319 & 0.0410 & 0.0416 & 0.0572 & 0.1143 & 0.0154 & 0.0333 \\
     & \textbf{DIETING} & 0.0638 & 0.1290 & 0.0410 & 0.0997 & 0.0562 & 0.1090 & 0.0154 & 0.0314 \\
     \midrule[1pt]
     \multirow{2}{*}{\textbf{Amazon-CD}} & \textbf{DIET} & 0.0425 & 0.0590 & 0.0410 & 0.1154 & 0.0356 & 0.0488 & 0.0154 & 0.0355 \\
     & \textbf{DIETING} & 0.0424 & 0.0589 & 0.0410 & 0.0780 & 0.0358 & 0.0489 & 0.0154 & 0.0294 \\
     \midrule[1pt]
     \multirow{2}{*}{\textbf{Amazon-TV}} & \textbf{DIET} & 0.0707 & 0.0896 & 0.0410 & 0.0764 & 0.0617 & 0.0771 & 0.0154 & 0.0488 \\
     & \textbf{DIETING} & 0.0705 & 0.0891 & 0.0410 & 0.0764 & 0.0620 & 0.0776 & 0.0154 & 0.0438 \\
     \bottomrule[2pt]
    \end{tabular}
}
\end{table}

\subsection{In-depth Analysis}
\subsubsection{Detail performance analysis}
\begin{figure}[htb]
  \centering
  \caption{The variation on the test set during training.}
  \label{fig:detail}
  \includegraphics[width=\linewidth]{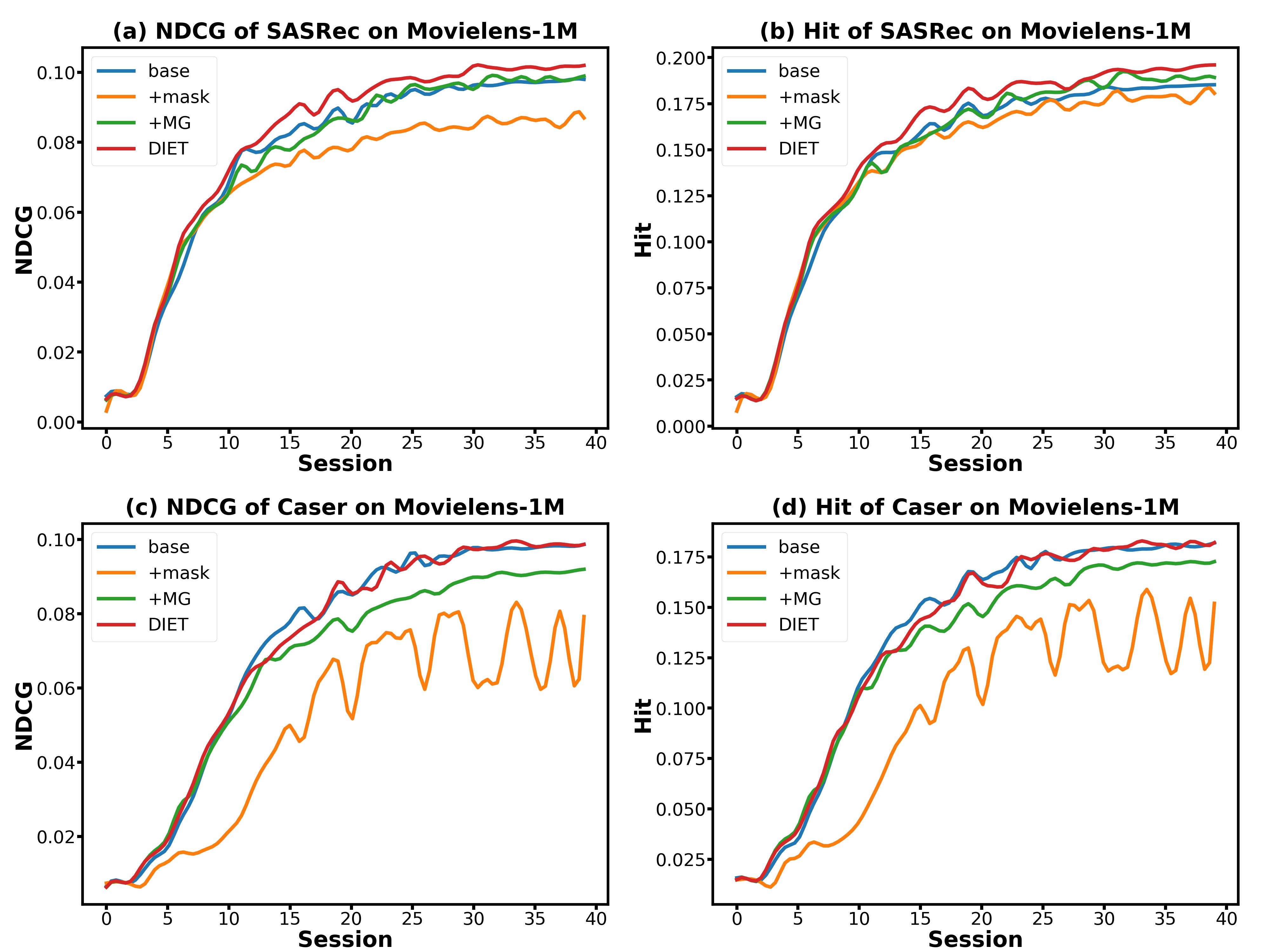}
\end{figure}

To further elucidate the effectiveness of our method, we plot the NDCG and Hit when training in Figure \ref{fig:detail}. It is evident that the personalized-based method outperforms \emph{mask}, indicating that personalization brings a huge performance improvement. The results also indicate that consistent diets for all users might lead to prediction oscillation, which is particularly severe on Caser. However, this figure also shows that simply assigning personalized diets might degrade the performance, such as Caser. In addition, our method with weight correction addresses this problem with better results than base model.

\subsubsection{Analysis on the sparsity ratio}

\begin{figure}[h]
  \centering
  \includegraphics[width=\linewidth]{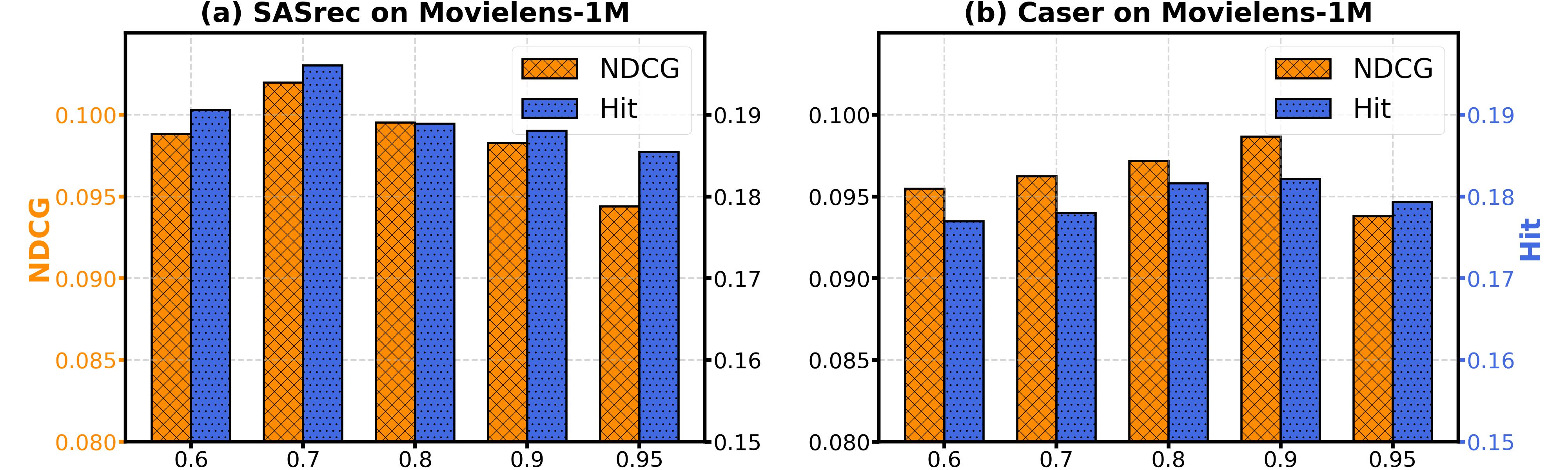}
  \caption{Sensitivity of each model towards $\alpha$.}
  \label{fig:sparsity}
\end{figure}

As the model on edges is frozen, the quantity of 0s and 1s in the mask sent from the cloud side determines the performance of the final model. Towards this end, we tune the sparsity of each model to gain insights into the potential of these subnets in incompatible networks. Denote $\alpha$ as the ratio of the number of elements equal to 0 in each layer to all elements. The range of $\alpha$ is $[0.6, 0.7, 0.8, 0.9, 0.95]$, and we plot the corresponding change of NDCG and Hit on Movielens-1M in Figure \ref{fig:sparsity}. On the one hand, when the value of $\alpha$ is small, then most of the mask is 1s and the mask learned is less effective. On the other hand, when $\alpha$ approaches 1, only a small fraction of elements can be selected, leading to a loss of model information, thereby decreasing the performance. From the figure we can clearly observe the performance of the two models on Movielens-1M first continues to increase as $\alpha$ grows, reaching a certain point after which it starts to decrease, for reasons as discussed earlier. Moreover, the results presented above confirm that properly finding specific $\alpha$ for each model is significant to balance the inference speed and the performance as more 0s lead to faster inference speed. 

\subsubsection{Analysis on the correction part}
\begin{figure}[htb]
  \centering
  \caption{Emperical study of the influence caused by connection correlation. We calculate the percent of non-zero rows/filters in each layer before and after correction.}
  \includegraphics[width=\linewidth]{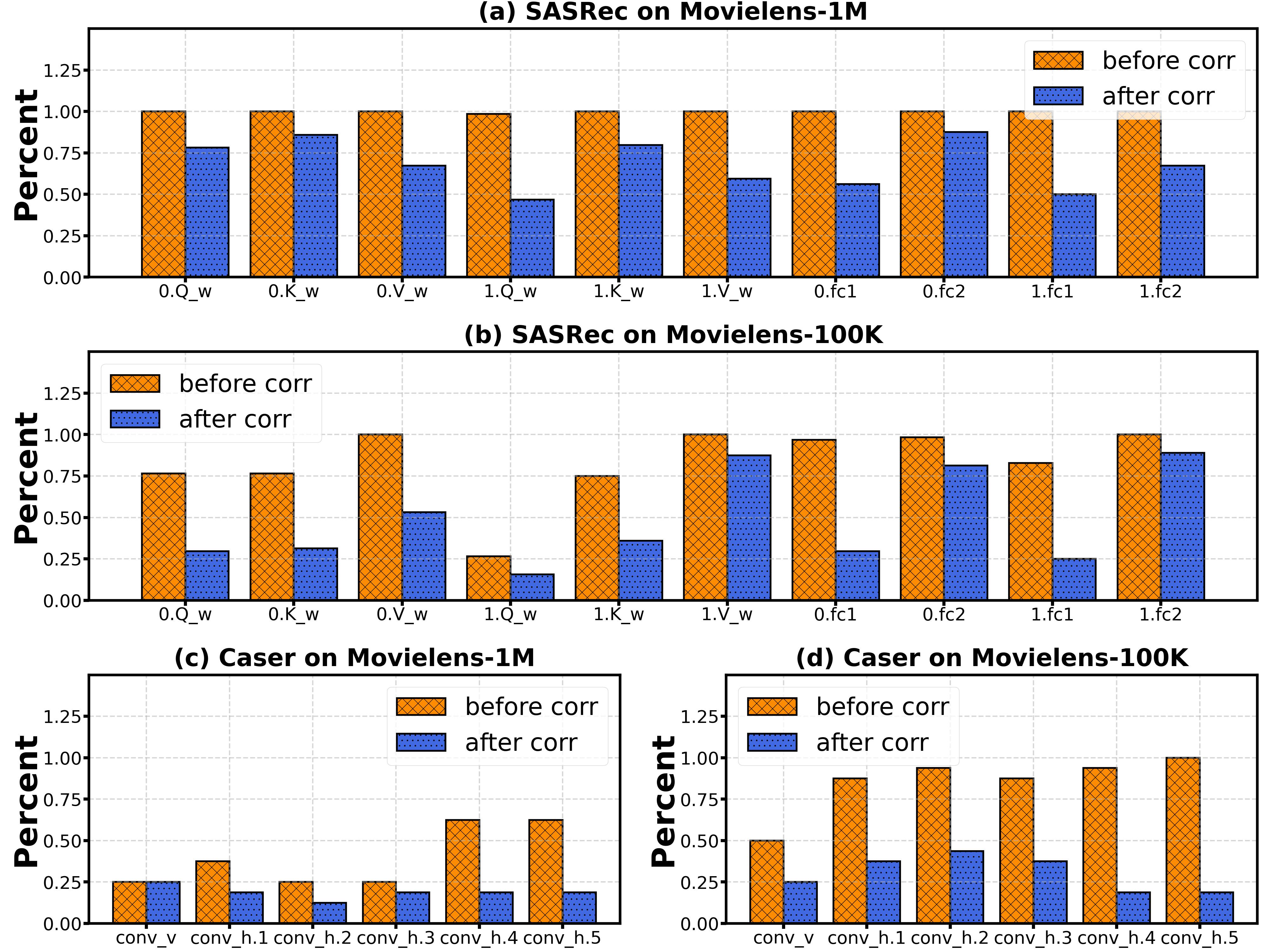}
  \label{fig:filter}
\end{figure}
In Section \ref{Ablation Study}, we have demonstrated that the weight correction reduces the number of floating-point operations and accelerates the inference speed of each edge. We deduce that this is due to the presence of more filters that consist entirely of zeros, which do not need to be involved in the inference process. As shown in figure \ref{fig:filter}, even with a small fraction of 1s in the mask, there are still a few filters that do not need to participate in calculation, especially in SASRec. This situation changes a little within Caser due to the small size of each filter in the convolutional layer. It suggests that reweighting each filter in each layer reduces the number of floating-point operations as assigning lower weights to less important filters reduces the scores of these elements in the final ranking, prompting the elements in these filters to become 0s. These results empirically verified the effectiveness of connections correction on both recommendation and rapid inference.

\subsubsection{Analysis on the generalizability}
\begin{table}[htb]
\caption{Performace comparison on two synthetic OOD datasets(Amazon-CD and Amazon-TV).} 
  \label{tab:experiment_generalizability}
  \centering
  \resizebox{1.0\columnwidth}{!}{
  
  \begin{tabular}{c|c|p{1.5cm}<{\centering}|p{1.5cm}<{\centering}|p{1.5cm}<{\centering}|p{1.5cm}<{\centering}}
    \toprule[2pt]
     \multirow{3}{*}{\textbf{Model}} & \multirow{3}{*}{\textbf{Method}} & \multicolumn{4}{c}{\textbf{Dataset}} \\
     \cline{3-6}
     & & \multicolumn{2}{c|}{\textbf{Amazon-CD}} & \multicolumn{2}{c}{\textbf{Amazon-TV}} \\
     \cline{3-6}
     & & \textbf{NDCG$@10$} & \textbf{Hit$@10$} & \textbf{NDCG$@10$} & \textbf{Hit$@10$} \\
     
    \midrule[1pt]
    \midrule[1pt]

    \multirow{4}{*}{\textbf{SASRec}} & \textbf{base} & 0.0360 & 0.0481 & 0.0747 & 0.0935 \\
    & \textbf{+mask} & 0.0390 & 0.0523 & 0.0722 & 0.0905\\
    & \textbf{+MG} & 0.0379 & 0.0503 & 0.0744 & 0.0930 \\
    & \textbf{DIET} & 0.0416 & 0.0577 & 0.0755 & 0.0946 \\

    \cline{1-6}
    
    \multirow{4}{*}{\textbf{Caser}} & \textbf{base} & 0.0309 & 0.0421 & 0.0585 & 0.0755 \\
    & \textbf{+mask} & 0.0114 & 0.0191 & 0.0304 & 0.0447\\
    & \textbf{+MG} & 0.0343 & 0.0461 & 0.0656 & 0.0820 \\
    & \textbf{DIET} & 0.0344 & 0.0461 & 0.0669 & 0.0831 \\
    
    \bottomrule[2pt]
  \end{tabular}
  }
\end{table}
Table \ref{tab:experiment_generalizability} shows the performance of our method on two synthetic datasets in order to prove the robustness of our score generator for those unseen user behavior sequences. Random $80\%$ users are used for training and the rest are for testing. We choose to use those Amazon datasets because of their large sparsity, this ensures that the sequence distribution of the training data and the testing data are inconsistent and helps justify whether models can learn accurate subnets from different user representations. From the table, we have the following findings: first, without personalized subnets for each edge, training a consistent model for all edges mostly yields the worst performance, especially for Caser on both datasets, where both metrics decreased by more than half. Second, those personalized models are consistently superior to it, demonstrating the high generalization capability of the hypernetwork condition on each user sequence.

\subsection{Case Study}
\begin{figure}[htb]
  \centering
  \includegraphics[width=\linewidth]{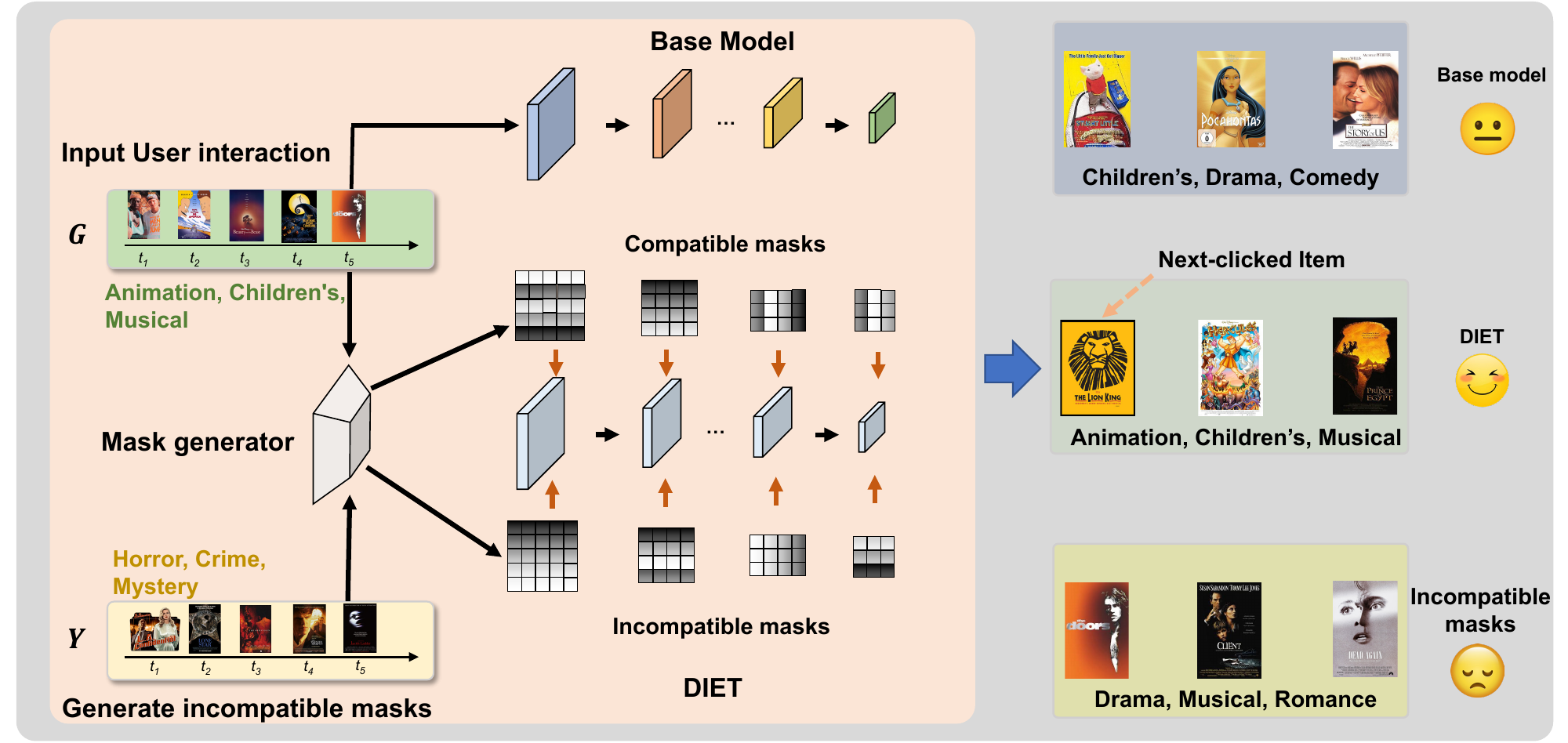}
  \caption{Case study for illustration of the compatible networks.}
  \label{fig:case_study}
\end{figure}

We present a case study in Figure \ref{fig:case_study} to illustrate the concept of compatibility as described in our paper. To demonstrate this, we extract two interactions from the MovieLens-1M dataset. The action denoted as \textbf{G}(the green one) is utilized to create compatible masks, while the action denoted as \textbf{Y}(the yellow one) is used to generate incompatible masks. Our findings reveal distinct preferences: \textbf{G} favors genres such as \emph{Animation, Children’s, and Musical}, whereas \textbf{Y} leans towards \emph{Horror, Crime, and Mystery}.

Initially, we input \textbf{G} into \emph{SASRec}, and observe that the top-3 recommended items do not include the next-clicked movies, but rather relevant movies. However, when employing compatible masks derived from \textbf{G} using DIET, the results significantly improve, with the next-clicked item and more relevant movies appearing in the recommended list. Conversely, when utilizing \textbf{Y} to generate masks, which exhibits disparate preferences compared to \textbf{G}, it recommends few relevant movies. Better performance compared to the trained \emph{SASRec} and inicompatible network highlights the superiority of DIET, underscoring that identifying compatible parameters can not only reduce resource costs but also yield superior results.

\section{Conclusion}
In this work, we investigate the problem of edge-cloud collaborative recommendation under communication costs and mobile edge resource constraints. In response to potential constraints during collaborative training, we have categorized them into three challenges and substantiated the necessity of addressing them. We propose an efficient framework named \textbf{DIET}, which takes both element and filter/row level importance into consideration and searches the most suitable diets for each edge, addressing the aforementioned challenges successfully. Based on it, we propose \textbf{DIETING} to utilize only one layer of parameters to represent the entire model but get comparable performance, thereby more storage-friendly. Further experiments on real-world datasets and widely used models, accompanied by insightful analyses, once again demonstrate the effectiveness of our framework.

\section*{Acknowledgement}
This work was supported by the National Science and Technology Major Project (2022ZD0119100), National Natural Science Foundation of China (62441605, 62376243, 62037001, U20A20387), Key Research and Development Program of Zhejiang Province(2024C03270), the StarryNight Science Fund of Zhejiang University Shanghai Institute for Advanced Study (SN-ZJU-SIAS-0010), Scientific Research Fund of Zhejiang Provincial Education Department (Y202353679).

\bibliographystyle{ACM-Reference-Format}
\balance
\bibliography{sample-base}

\appendix
\setcounter{table}{0} 

\section{Evaluation Metrics}
For the edge-cloud collaborative learning in recommendation, we are supposed to consider recommender performance, transmission latency and inference speed simultaneously. For the recommender performance, we use Hit and NDCG, two frequently used metrics in recommender system . The first metric assesses whether the target items appear in the recommender’s provided recommendation list. This is equivalent to \emph{recall} when the number of target items is 1, as in our experiments. The second metric is used to measure the quality of recommendation lists and the accuracy of their ordering, it combines the relevance of recommended items with the impact of their ranking. Detailed calculation process is as follows: 
\begin{itemize}
    \item \textbf{NDCG} takes into account both the relevance of the items and their positions in the ranked list. It considers that highly relevant items appearing at the top of the list are more valuable to the user:
            \begin{equation}
                NDCG@N = \frac{DCG@N}{IDCG@N},
            \end{equation}
            where the numerator $DCG@N=\sum_{i=0}^{i=N}\frac{r_i}{log_2(i+1)}$ and the denominator $IDCG@N=\sum_{i=0}^{i=\mathop{\min}(N, K)}\frac{1}{log_2(i+1)}$. $N$ and $K$ in the above equations are the customized length of items on top of the recommender list and the length of the user interaction list for evaluation (in our experiment the last interaction items). $r_i$ in the first equation denotes whether item $i$ in the recommender list is in the user interaction list for evaluation.
    \item \textbf{Hit Rate} represents whether there exists any item in our recommendation list:
    \begin{equation}
        Hit@N = \sum_{i=0}^{i=N}r_i > 0.
    \end{equation}
\end{itemize}
For the transmission latency, we evaluate it with the number of bits needed to transfer when updating. For the base models like \textit{SASRec} and \textit{Caser}, the transmission cost is calculated as $32\times N_p$, where $N_p$ is the number of parameters in the models. Aligned with the popular pruning settings, compressed sparse column (CSC) or compressed sparse row (CSR) formats are used to store the parameters for most baselines. They are storage formats for sparse matrices, efficiently designed for column-oriented operations by storing values, column/row indices, and offsets. Transmission cost of them is calculated as $32\times 2 \times \alpha \times N_p$. $\alpha$ in the equation is the proportion of non-zero elements in the mask of each layer. For \textit{DIET}, \textit{DIETING}, and those sparse-selection methods, we use binary format to store the masks, which is $N_p$. For the inference speed, we use Floating-point operations(FLOPs) of a single iteration during inference. 
\begin{table}[htb]
\caption{Performace comparison between DIET and DIETING on two synthetic OOD datasets.} 
\label{tab:appendix_generalizability}
\centering
\resizebox{1.0\columnwidth}{!}{

    \begin{tabular}{c|c|p{1.5cm}<{\centering}|p{1.5cm}<{\centering}|p{1.5cm}<{\centering}|p{1.5cm}<{\centering}}
    \toprule[2pt]
     \multirow{3}{*}{\textbf{Model}} & \multirow{3}{*}{\textbf{Method}} & \multicolumn{4}{c}{\textbf{Dataset}} \\
     \cline{3-6}
     & & \multicolumn{2}{c|}{\textbf{Amazon-CD}} & \multicolumn{2}{c}{\textbf{Amazon-TV}} \\
     \cline{3-6}
     & & \textbf{NDCG$@10$} & \textbf{Hit$@10$} & \textbf{NDCG$@10$} & \textbf{Hit$@10$} \\
     
    \midrule[1pt]
    \midrule[1pt]
    
    \multirow{2}{*}{\textbf{SASRec}} & \textbf{DIET} & 0.0416 & 0.0577 & 0.0755 & 0.0946 \\
    & \textbf{DIETING} & 0.0411 & 0.0573 & 0.0748 & 0.0937 \\
    
    \cline{1-6}
    
    \multirow{2}{*}{\textbf{Caser}} & \textbf{DIET} & 0.0344 & 0.0461 & 0.0669 & 0.0831 \\
    & \textbf{DIETING} & 0.0345 & 0.0469 & 0.0665 & 0.0832 \\
    
    \bottomrule[2pt]
    \end{tabular}
  }
\end{table}

\section{Implementation Details}
Due to the potential inconsistency in performance of different pre-trained models when searching for recipes, we therefore use random initialized parameters in our experiments to get rid of the unfairness brought by different pre-trained model. For all the models mentioned above, we use Adam as the optimizer with a triangular learning rate scheduler. The learning rate is 0.001 for \emph{Movielens-100K} and 0.1 for others. The size item embedding we use in our experiment is 64. We use Xavier normal initialization for all the model parameters. The number of horizontal and vertical convolution filters are 4 and 16 in \textit{Caser}, respectively. For the activation functions $\phi_a$ and $\phi_c$, we simply use tanh in our experiments. For \textit{SASRec} we use two self-attention blocks and four heads in each block. The dropout rate is set to zero and the max length of each input interaction is 5. The hyper-parameter $\alpha$ across datasets and models is significantly different. In \textit{Caser} $\alpha$ is set to 0.1, 0.1, 0.2, 0.1 on Movielens-1M, Movielens-100K, Amazon-CD and Amazon-Movies, while in \textit{SASRec} $\alpha$ is set to 0.3, 0.1, 0.1, 0.1 respectively. As none of SuperMask and Gater have a precise way to control the model sparsity, we try to modify the initialized value and the coefficient of the penalty term as much as possible.

\section{generalizability of DIETING}
In order to further investigate the feasibility of using one layer to obtain all the models, we test it on those synthetic out-of-distribution datasets, and results are shown in Table \ref{tab:appendix_generalizability}. From the table, we can observe that with personalized mask generator and inter-layer corrections, storing one layer on edges still keeps reliable generalizability.

\end{document}